%% file: main.tex
\documentclass[11pt,a4paper,twocolumn]{huawei-techreport}

\usepackage[authoryear,sort&compress,round]{natbib}
\usepackage{xurl}
\usepackage{multicol}
\input{sections/appendix_prompt_style}
\usepackage{CJKutf8}
\AtBeginDocument{\begin{CJK}{UTF8}{gbsn}}
\AtEndDocument{\end{CJK}}
\input{paper-title}
\title{\papertitle}
\renewcommand\Affilfont{\normalfont\fontsize{10}{12}\selectfont}
\author{Alan Tang\textsuperscript{1}, Lifeng Wang\textsuperscript{1,2}, Gang Wu\textsuperscript{1}, Tao Zhang\textsuperscript{1}, Yantao Jia\textsuperscript{1,*}, and Xiaofeng Zhu\textsuperscript{1}}
\affil{\textsuperscript{1}Huawei Technologies Co., Ltd.\quad \textsuperscript{2}Harbin Institute of Technology, Shenzhen}
\usepackage{etoolbox}
\patchcmd{\maketitle}{\vskip11pt}{\vskip6pt}{}{\PackageError{skillmark}{Title spacing patch failed}{}}
\patchcmd{\maketitle}{\vskip20pt}{\vskip10pt}{}{\PackageError{skillmark}{Author spacing patch failed}{}}
\patchcmd{\maketitle}{\vskip20pt}{\vskip4.5pt}{}{\PackageError{skillmark}{Abstract spacing patch failed}{}}
\authfootnote{\textsuperscript{*}Corresponding author.}
\renewcommand{\today}{October 10, 2026}
\hypersetup{
  pdftitle={\papertitle},
  pdfauthor={Alan Tang, Lifeng Wang, Gang Wu, Tao Zhang, Yantao Jia, Xiaofeng Zhu},
  pdfsubject={SkillMarker manuscript draft}
}

\input{sections/abstract}

\begin{document}
\maketitle

\input{paper-body}

\renewcommand{\bibfont}{\small}
\setlength{\bibsep}{2pt}
\onecolumn
\begin{multicols}{2}
\bibliography{references}
\end{multicols}

\clearpage
\appendix
\input{sections/appendix}
\input{sections/arr/ablation}

\end{document}

%% file: sections/appendix_prompt_style.tex
\usepackage{tcolorbox}
\tcbuselibrary{breakable}
\usepackage{fvextra}

\newtcolorbox{appendixpromptbox}[1]{
  title={#1},
  title after break={#1\space (continued)},
  fonttitle=\small\bfseries,
  boxrule=2pt,
  arc=0mm,
  colback=black!5!white,
  colframe=black!75!white,
  breakable,
  before skip=10pt,
  after skip=10pt,
  pad at break=1mm
}

\ExplSyntaxOn
\cs_new_protected:Npn \AppendixPromptFormatText #1
  {
    \regex_match:nnTF { \A \( Omitted: } {#1}
      { \textbf{#1} }
      { #1 }
  }
\ExplSyntaxOff

\DefineVerbatimEnvironment{appendixprompttext}{Verbatim}{
  fontsize=\scriptsize,baselinestretch=0.98,breaklines=true,breaksymbolleft={},
  formatcom=,
  frame=lines,framesep=2mm,tabsize=4
}

\DefineVerbatimEnvironment{appendixteachertext}{Verbatim}{
  fontsize=\scriptsize,baselinestretch=0.98,breaklines=true,breaksymbolleft={},
  formatcom=,
  frame=lines,framesep=2mm,tabsize=4,commandchars=+\[\]
}

%% file: paper-title.tex
\newcommand{\papertitle}{SkillMarker: A KGW-Based Watermarking Method for Agent Skill Ownership Verification}

%% file: sections/abstract.tex
\begin{abstract}
Agent skills encode operational procedures and domain expertise as reusable knowledge assets, yet their distribution in plaintext exposes them to unauthorized copying and leakage. However, practical mechanisms for skill ownership verification and leakage tracing remain underdeveloped. Watermark embedding before distribution offers a potential solution, but structured, compact skill instructions impose stringent requirements for factual and procedural accuracy. To address these challenges, we introduce SkillMarker, a KGW-based watermarking method that jointly targets functionality preservation and strong watermark embedding. SkillMarker combines quality-constrained rewriting with signal-oriented candidate search to strengthen watermark signals while preserving essential instructions.  We develop a toolkit\ for watermark embedding, quality auditing, and detection. Across four datasets, SkillMarker negligibly affects downstream task performance on average. In our copy-paste tests, window-based detection consistently identifies copied passages containing as few as 1,536 consecutive tokens. Even after merging with one or three background skills, SkillMarker retains watermark detection rates of 72.2--77.8\%.  Project homepage: \url{https://fuxi-ailabs.github.io/SkillMarker/}.
\end{abstract}

%% file: paper-body.tex
\input{sections/introduction}

\input{sections/method}

\input{sections/experiments}
\input{sections/related_work}
\input{sections/conclusion}

%% file: sections/introduction.tex
\section{Introduction}
\label{sec:introduction}

Agent skills package domain expertise into reusable instructions and resources that agents load on demand to specialize their behavior without modifying model weights~\citep{anthropic2025agentskills,li2026skillsbench}. This encoded knowledge makes skills valuable assets for organizations and developers, while their reusable format allows that expertise to be deployed across tasks. However, distributing skills in plaintext makes them easy to copy, redistribute, or incorporate into other skills.

\begin{figure}[!t]
    \centering
    \captionsetup{font=normalsize,skip=4pt}
    
    \includegraphics[width=0.90\linewidth,height=0.60\textheight,keepaspectratio]{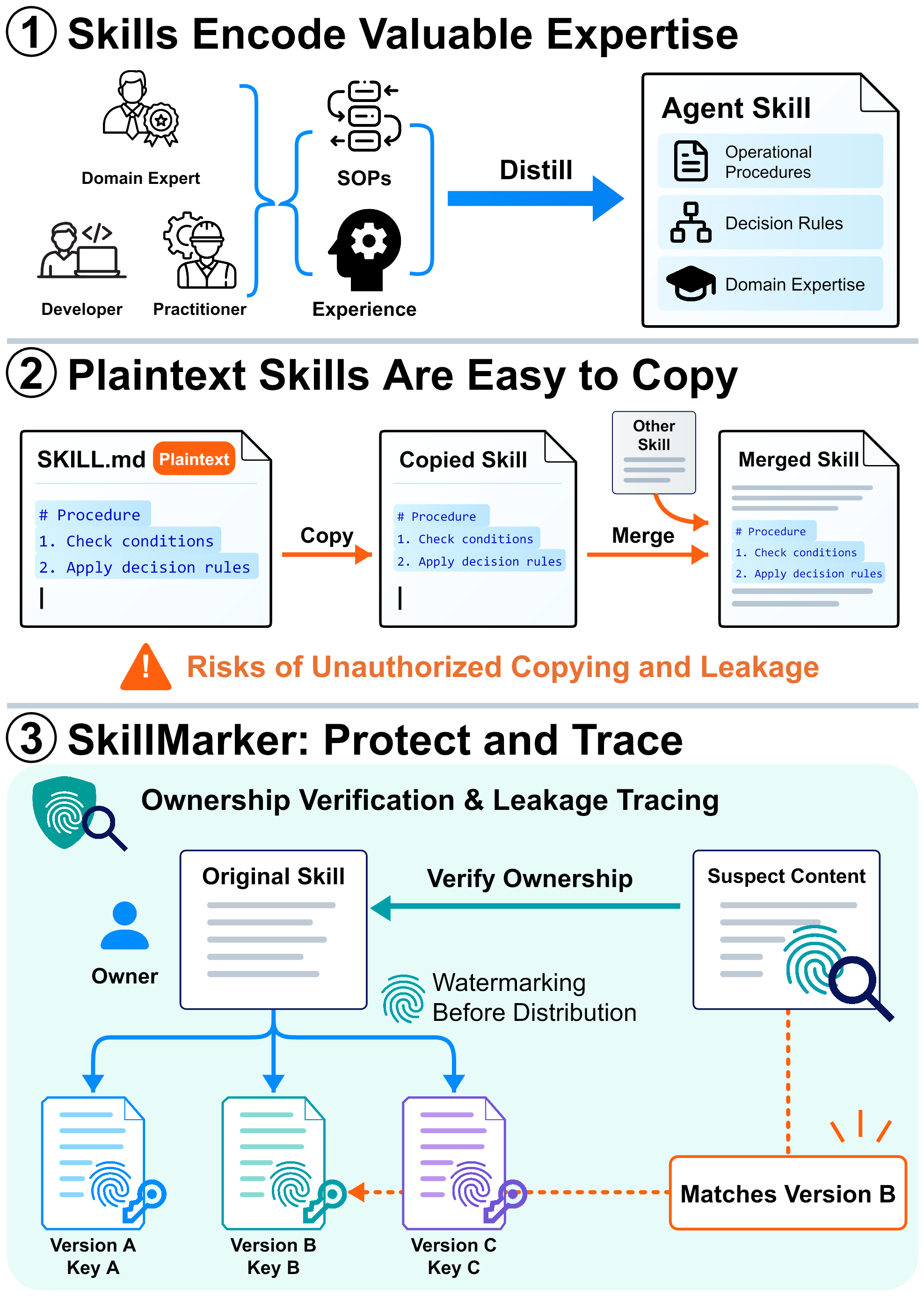}
    
    \vspace{0pt}
    \caption{Plaintext skills expose valuable domain expertise to unauthorized copying and leakage. \textbf{SkillMarker} offers ownership verification and leakage tracing.}
    \label{fig:skillmark_motivation}
    \vspace{-2pt}
\end{figure}

These risks motivate watermarking skill text before distribution to support ownership verification and leakage tracing, including when only fragments of the original skill remain. We therefore ask: \textbf{Can we watermark the skill for ownership verification and leakage tracing without compromising its functionality?}

However, embedding a watermark into structured, compact skill instructions with stringent accuracy requirements poses two coupled challenges. \textbf{C1: Preserving skill functionality during watermarking.} Skill instructions encode precise operational requirements, so even minor changes to resource references, decision conditions, or required actions can change agent behavior. Their structured format and stringent accuracy requirements leave limited room for watermark embedding through rewriting. \textbf{C2: Embedding a strong watermark under limited rewriting freedom.} Compact skill instructions offer few opportunities to strengthen the watermark signal. Accuracy constraints further restrict alternative expressions: a weak logit bias may produce little watermark signal, while a stronger bias can change essential instructions.

We introduce \textbf{SkillMarker}, a KGW-based watermarking method for agent skill ownership verification~\citep{kirchenbauer2023watermark}. To address C1, we propose \emph{quality-constrained rewriting}, which preserves protected content and constrains editable spans to preserve operational requirements. To address C2, we propose \emph{signal-oriented candidate search}, which explores alternative expressions and selects candidates with stronger watermark signals among those that pass preservation checks.

SkillMarker enables developers to \textbf{verify skill ownership} by embedding a keyed watermark before publishing a skill and later testing suspect text for a watermark signal. It also enables them to \textbf{trace skill leakage} by assigning distinct keys to distributed versions and testing suspect text for a watermark signal that can help identify its source version, including when instructions are copied in whole or incorporated into other skills. To support these applications, we also develop a toolkit that integrates watermark embedding, quality auditing, and detection.

Our contributions are:
\begin{itemize}[leftmargin=0pt,itemindent=0pt,listparindent=0pt,labelsep=0.5em]
    \item We propose SkillMarker for skill ownership verification and leakage tracing. By adapting KGW through quality-constrained rewriting and signal-oriented candidate search, SkillMarker addresses two challenges: preserving skill functionality and embedding strong watermarks under limited rewriting freedom.
    \item Across four datasets spanning general and professional tasks, we show that watermarked skills maintain task performance with negligible degradation on average compared with original skills.
    \item We evaluate watermark robustness under two common skill-reuse scenarios: copy-paste and skill merging. Our detection method consistently identifies copied passages containing as few as 1,536 consecutive tokens, while detection rates remain at 72.2--77.8\% after merging with multiple background skills.
\end{itemize}

%% file: sections/method.tex
\section{Method}
\label{sec:method}

\begin{figure*}[!t]
    \centering
    \includegraphics[width=\textwidth,height=0.28\textheight,keepaspectratio]{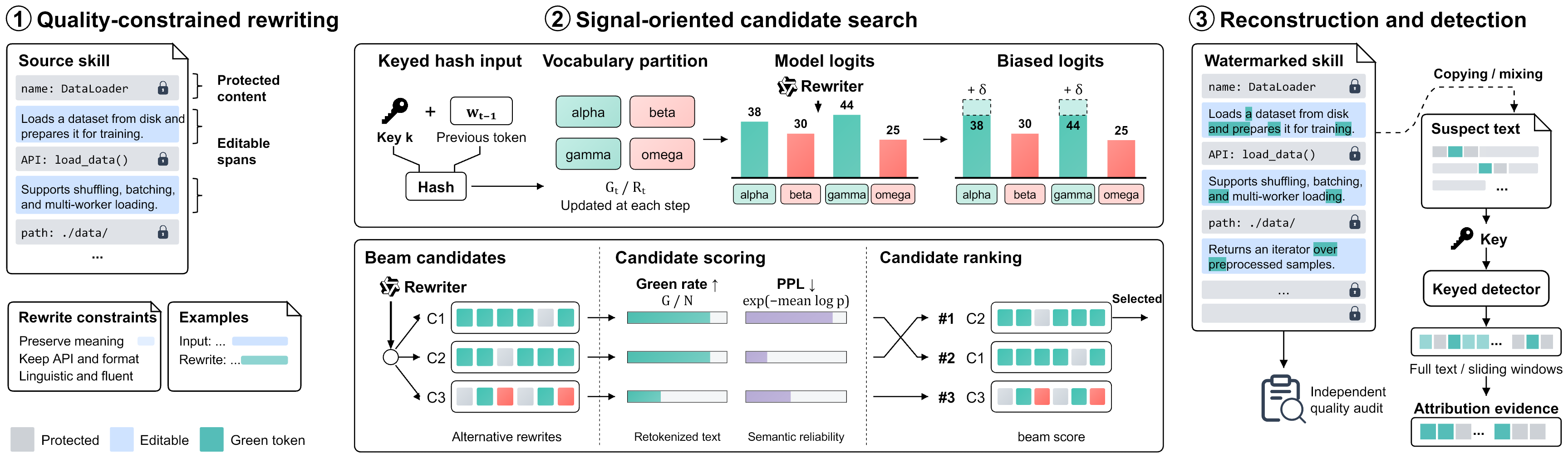}
    \caption{Overview of SkillMarker. \textbf{(1)} Quality-constrained rewriting preserves protected content, such as code and API identifiers, while rewriting editable spans. \textbf{(2)} Signal-oriented candidate search ranks candidate rewrites by the green-token rate and uses lower perplexity to break ties. \textbf{(3)} Reconstruction preserves protected content, followed by independent quality auditing and detection in suspect text.}
    \label{fig:skillmark_method}
\end{figure*}

\newcommand{\skillmarktoken}[1]{\begingroup\setlength{\fboxsep}{1.5pt}\setlength{\fboxrule}{0.3pt}\fcolorbox{black!35}{black!5}{\texttt{\small\strut #1}}\endgroup}

\subsection{Problem Setup and Overview}
\label{sec:method_setup}

Let $x$ denote the Markdown body of an original skill. We introduce a watermark key $k$ for watermark embedding and detection. We seek a watermarked skill $\widetilde{x}$ that preserves the instructions in $x$ while carrying a watermark signal associated with $k$. A detector later tests suspect text for this signal, including when only part of the distributed skill is present. Watermark embedding uses a language model with access to its output logits; detection requires the tokenizer and watermark configuration, but no model inference.

For a vocabulary $\mathcal{V}$, write a token sequence as $w_{1:n}$. At generation step $t$, the input context and preceding generated tokens determine a next-token distribution $P_t$ through a softmax over the model's output logits. The watermark key $k$ and preceding token $w_{t-1}$ determine a green list $G_t$ and its complement, the red list $R_t=\mathcal{V}\setminus G_t$; $\gamma\in(0,1)$ controls the nominal green-list fraction, and $\delta>0$ controls the logit bias. We use $Q_t$ for the resulting watermarked distribution. The set $E\subset\mathcal{V}$ contains tokens exempt from logit bias and scoring. For any scored text, $\mathcal{I}$ denotes its eligible token positions, $N=|\mathcal{I}|$ their count, and $G$ counts green tokens.

Figure~\ref{fig:skillmark_method} provides an overview of SkillMarker. The central difficulty is to strengthen watermark signals within the limited rewriting freedom allowed by precise skill instructions. We first separate protected content from editable spans, then use preservation instructions and paired rewrite examples to guide faithful reformulation. Within each editable span, beam search explores alternative expressions under the KGW-biased distribution, seeking candidates that carry a stronger watermark signal while retaining the original operational requirements. We assess watermark signal strength using the green-token rate and linguistic fluency using perplexity (PPL)~\citep{bengio2003neural}. Among candidates that pass preservation checks, we select those with higher green-token rates, using lower PPL to break ties. We then reconstruct the watermarked skill with protected content unchanged. The toolkit provides independent quality auditing and supports both full-text and window-based watermark detection.

\subsection{Quality-Constrained Rewriting}
\label{sec:quality_rewriting}

\paragraph{Protecting literal content.}
To embed a watermark, we apply a logit bias that favors green-list tokens during rewriting. Without preservation constraints, rewriting can change executable code and invalidate a skill. For example, changing \skillmarktoken{>=} to \skillmarktoken{>} in the following code block changes when the retry limit is enforced:
\begin{mdframed}[backgroundcolor=black!5,linecolor=black!35,linewidth=0.3pt,skipabove=5pt,skipbelow=5pt,innerleftmargin=6pt,innerrightmargin=6pt,innertopmargin=6pt,innerbottommargin=6pt,nobreak=true]
\small
\begin{verbatim}
if attempts >= 3:
    raise RuntimeError
\end{verbatim}
\end{mdframed}
Code and numerical thresholds allow little rewriting freedom, whereas surrounding prose often allows alternative expressions. We therefore parse the original skill and designate code, structural delimiters, and recognized literals as protected content that must remain unchanged. Regions with unresolved structure are also protected. We split the remaining prose into editable spans and protect tokens that cross the boundary of protected content. Editable spans below a threshold on the original token count remain unchanged. We rewrite one editable span at a time and then reconstruct the skill with protected content in its original order.

\paragraph{Preserving instruction fidelity.}
Preserving protected content alone does not preserve a decision rule: changing \skillmarktoken{all conditions} to \skillmarktoken{any condition} changes behavior even if every threshold remains unchanged. We use in-context learning to make preservation constraints within editable spans concrete. The rewriting prompt combines explicit preservation instructions with paired input and rewrite examples. These examples show how to change wording while preserving the original language, facts, conditions, negations, numerical values, and technical identifiers. Preservation checks reject incomplete or empty candidates, unauthorized layout or syntax changes, and changes to protected content. Independent quality auditing then assesses language, semantics, completeness, and usability without exposing watermark scores to the reviewers. Downstream task performance is evaluated separately.

\paragraph{Grouping token variants.}
Watermark embedding can favor token variants that disrupt formatting. For example, a hard watermark that blocks \skillmarktoken{name} but permits \skillmarktoken{\_name} can produce an unwanted underscore and change Markdown formatting. A logit bias can likewise favor a token variant without excluding the original token. We therefore assign consistent green-list membership to token variants. Our implementation groups tokens that differ only in leading spaces or periods and assigns them a shared representative's green-list membership; literal-underscore variants remain subject to preservation checks. Specifically, we sample a keyed subset of $\lfloor\gamma|\mathcal{V}|\rfloor$ vocabulary IDs and assign each representative's membership to all tokens in its group. The resulting green-list size need not equal $\gamma|\mathcal{V}|$, so $\gamma$ remains a nominal parameter.

\paragraph{Exempting formatting tokens.}
Applying a logit bias to Markdown symbols can favor formatting changes that disrupt skill structure. We therefore exclude designated formatting tokens from the logit bias. Together with special and reserved control tokens, they form $E$. Following soft KGW~\citep{kirchenbauer2023watermark}, we modify the remaining logits by
\begin{equation}
    Q_t(v)=\frac{P_t(v)\exp\!\left(\delta\mathbf{1}\{v\in G_t\setminus E\}\right)}
    {\sum_{u\in\mathcal{V}}P_t(u)\exp\!\left(\delta\mathbf{1}\{u\in G_t\setminus E\}\right)}.
    \label{eq:skillmark_bias}
\end{equation}
Tokens in $R_t$ remain available, allowing the model to select strongly preferred expressions. Exempt tokens receive no logit bias and contribute no watermark signal.

\subsection{Signal-Oriented Candidate Search}
\label{sec:signal_search}

\paragraph{Exploring alternative expressions.}
Strict preservation constraints can encourage copying the original text without changes, leaving little room for additional watermark signal. Increasing $\delta$ alone may instead favor an inaccurate expression. We therefore use beam search~\citep{sutskever2014sequence} to explore alternative expressions under the same preservation constraints. It selects likely continuations under $Q_t$ in Eq.~\eqref{eq:skillmark_bias}, producing completed candidates for each editable span.

\paragraph{Selecting on reconstructed text.}
A candidate's generated token sequence can change when its text is joined to adjacent content. Scoring that sequence alone may therefore favor a watermark signal absent from the watermarked skill. We insert each candidate into the current skill and tokenize the complete reconstructed text. For the current editable span, $\mathcal{I}$ contains only nonexempt tokens wholly inside its output boundaries and having a preceding token. Each token's green-list membership is determined using its actual preceding token, even when that token is protected or exempt. The candidate's green-token rate is
\begin{equation}
    r=\frac{G}{N},\qquad
    G=\sum_{t\in\mathcal{I}}\mathbf{1}\{w_t\in G_t\}.
    \label{eq:skillmark_rate}
\end{equation}
We compute conditional PPL over each candidate's generated tokens under the rewriting prompt to assess linguistic fluency. PPL is the exponential of the mean token negative log-likelihood under $P_t$.

For editable span $i$, let $\mathcal{C}_i$ contain candidates that pass the preservation checks and have $N>0$. We select
\begin{equation}
    C_i^*\in\operatorname*{arg\,max}_{C\in\mathcal{C}_i}
    \bigl(r(C),-\mathrm{PPL}(C)\bigr),
    \label{eq:skillmark_selection}
\end{equation}
using lexicographic order: higher green-token rates take priority, and lower PPL breaks ties. If $\mathcal{C}_i$ is empty, we preserve the original editable span and record a fallback. We process editable spans in their original order and score the saved watermarked skill after reconstruction. This local selection does not guarantee the highest green-token rate for the complete skill.

\subsection{Watermark Detection}
\label{sec:watermark_detection}

Suspect text may contain copied instructions mixed with unrelated content, and the positions of protected content may be unavailable. We therefore support both full-text and window-based detection using only the suspect text and watermark configuration. Window-based detection uses sliding windows to detect watermark signals within suspect text. The detector tokenizes the full suspect text once and reconstructs $G_t$ from $k$ and each observed preceding token, using the same token-variant grouping as watermark embedding. It excludes $E$ and the first token. Without the positions of protected content, all remaining positions are eligible, including content that may have been protected during watermark embedding.

For $N>0$, we report the green-token rate $G/N$ and the KGW statistic
\begin{equation}
    z=\frac{G-\gamma N}{\sqrt{N\gamma(1-\gamma)}}.
    \label{eq:skillmark_detection}
\end{equation}
The detector computes the nominal upper-tail probability $\Pr[\operatorname{Binomial}(N,\gamma)\geq G]$. For window-based detection, it scores overlapping windows of the same tokenization, preserving each token's preceding token in the full suspect text, and applies a Bonferroni correction across the scanned windows. Repeated contexts and shared green-list membership can violate the binomial null assumptions; these probabilities therefore require independent calibration before being interpreted as empirical false-positive guarantees.

%% file: sections/experiments.tex
\begin{figure*}[!t]
    \centering
    \includegraphics[width=0.72\textwidth]{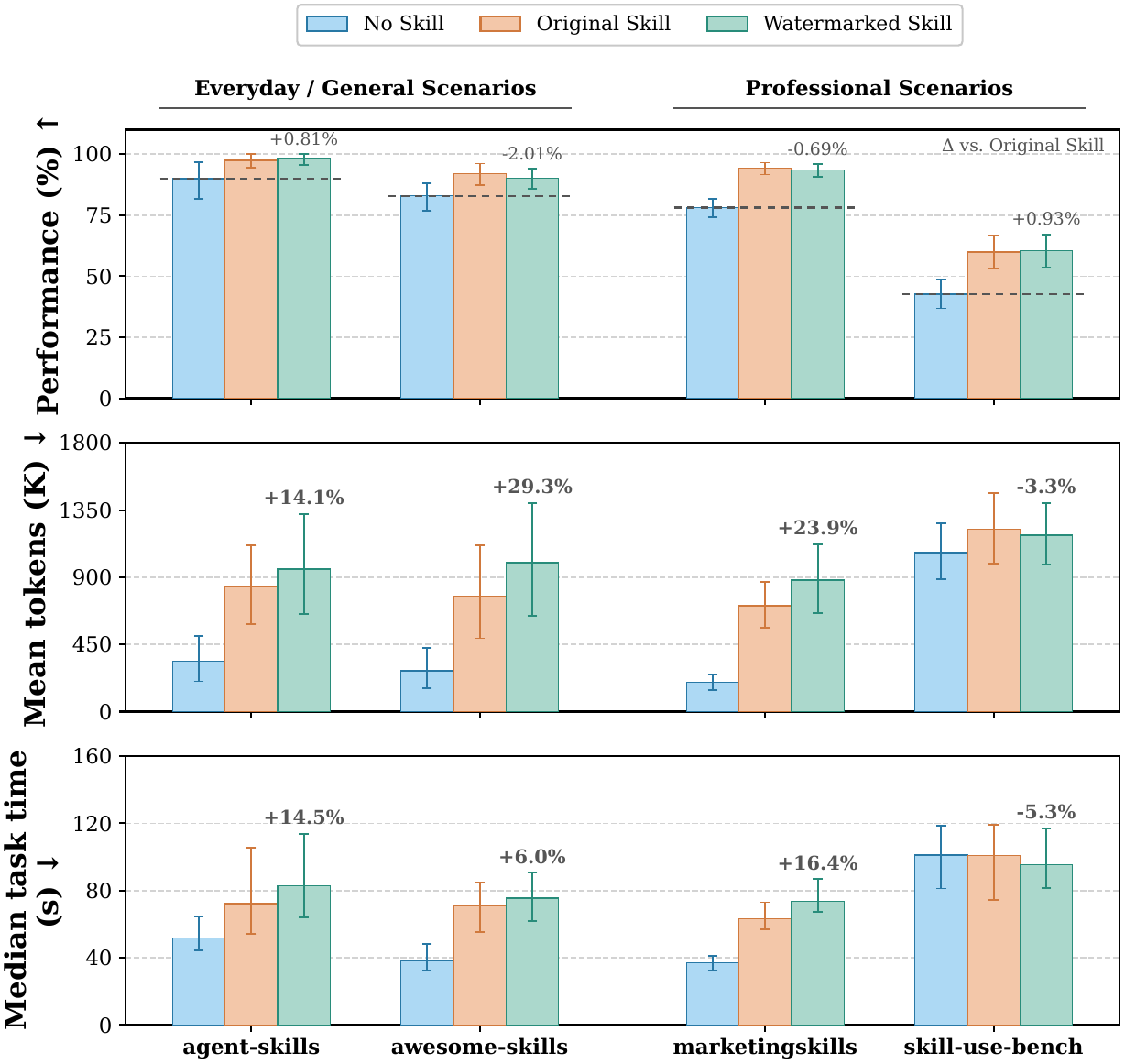}
    \caption{\textbf{Watermarking largely preserves downstream task performance.} We test watermarked skills on four datasets, comparing task performance, token usage, and task time against original skills and a no-skill baseline.}
    \label{fig:downstream_results}
\end{figure*}

\section{Experiments}
\label{sec:experiments}

\subsection{Experimental Setup}
\label{sec:experimental_setup}

We evaluate SkillMarker on four datasets spanning everyday/general
and professional scenarios. For everyday/general scenarios, \texttt{agent-skills}
\citep{osmani2026agentskills} covers routine software engineering workflows,
including planning, code review, testing, and debugging, while
\texttt{awesome-skills} \citep{sickn332026awesomeskills} provides a broader
collection for writing, document processing, information retrieval, and data
analysis. For professional scenarios, \texttt{marketingskills}
\citep{haines2026marketingskills} focuses on marketing workflows such as content
strategy, search engine optimization, advertising, and conversion optimization;
\texttt{skill-use-bench} \citep{han2026skilluse} provides executable tasks
requiring specialized procedures, including software development, document
processing, and scientific computation. 
We use Qwen3-8B~\citep{yang2025qwen3} for watermark embedding. We compare three settings: no skill, the original skill, and the watermarked skill.
We use DeepSeek-V4-Flash through Open Code, with one
DeepSeek-V4-Flash judgment per case.
Each case runs in a separate working directory with only the target-skill version required by its setting; the no-skill baseline receives neither version. Expected outputs and grading materials are withheld from the executing agent to prevent evaluation leakage.

\subsection{Downstream Performance and Efficiency}
\label{sec:downstream_evaluation}

Figure~\ref{fig:downstream_results} reports task performance (the percentage of evaluation checks passed), mean
total token usage, and median task time for no skill, original skills, and watermarked skills.
Token counts include input, cached input, and output (K = 1000), and bar labels report relative changes from original skills.

Watermarking largely preserves downstream task performance: watermarked skills
change performance by $-2.01$\% to $+0.93$\% relative to original skills
across four datasets and retain their gains over the no-skill baseline.
On \texttt{agent-skills}, pass rates are 97.46\% before watermarking and
98.25\% afterward; on \texttt{skill-use-bench}, they are 59.85\% and 60.41\%.
Original skills improve performance over no-skill runs by 8.49--40.30\%.
Mean total tokens increase by 14.1--29.3\% on the first three datasets but
decrease by 3.3\% on \texttt{skill-use-bench}. These means are sensitive to
the long-tailed distribution of run token counts.
Median task time increases by 14.5\%, 6.0\%, and 16.4\% for
\texttt{agent-skills}, \texttt{awesome-skills}, and \texttt{marketingskills},
respectively, and decreases by 5.3\% for \texttt{skill-use-bench}.
Task time is reported for reference: service load, execution periods, and
concurrency can still affect these measurements. Our assessment of
functionality preservation relies on task performance rather than timing differences.
The values are provided in Table~\ref{tab:skillmark_main_results}
in Appendix~\ref{app:downstream_results}.

\begin{figure*}[!t]
    \centering
    \includegraphics[width=0.72\textwidth]{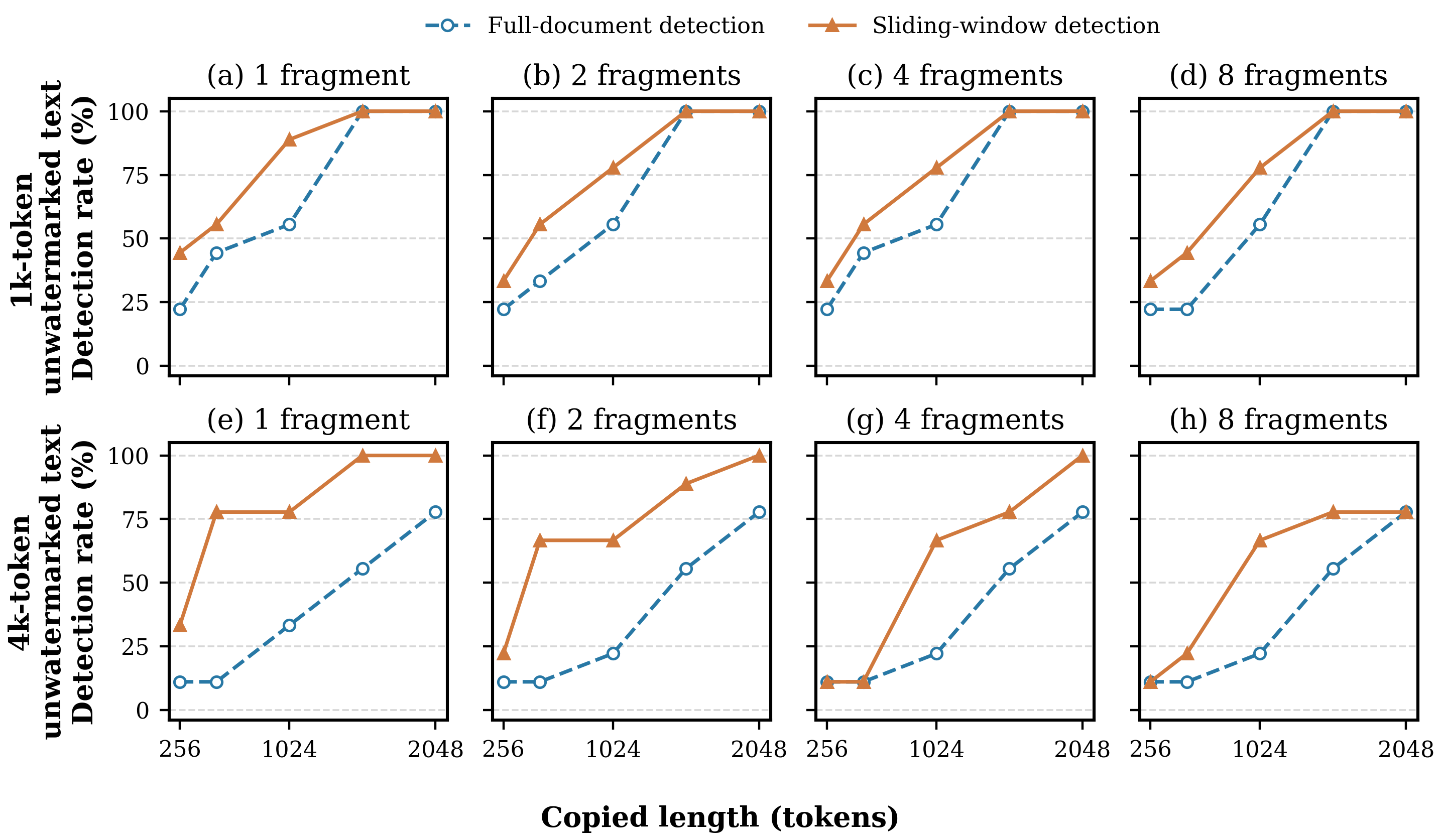}
    \caption{\textbf{Copied passages containing as few as 1,536 consecutive tokens are consistently detected in our tests.} We insert watermarked passages into unwatermarked text and compare full-text and window-based detection. Window-based detection identifies every tested contiguous 1,536-token copy under both background lengths.}
    \label{fig:copy_paste_detection}
\end{figure*}

\begin{figure*}[!t]
    \centering
    \includegraphics[width=0.72\textwidth,trim=0 100bp 0 0,clip]{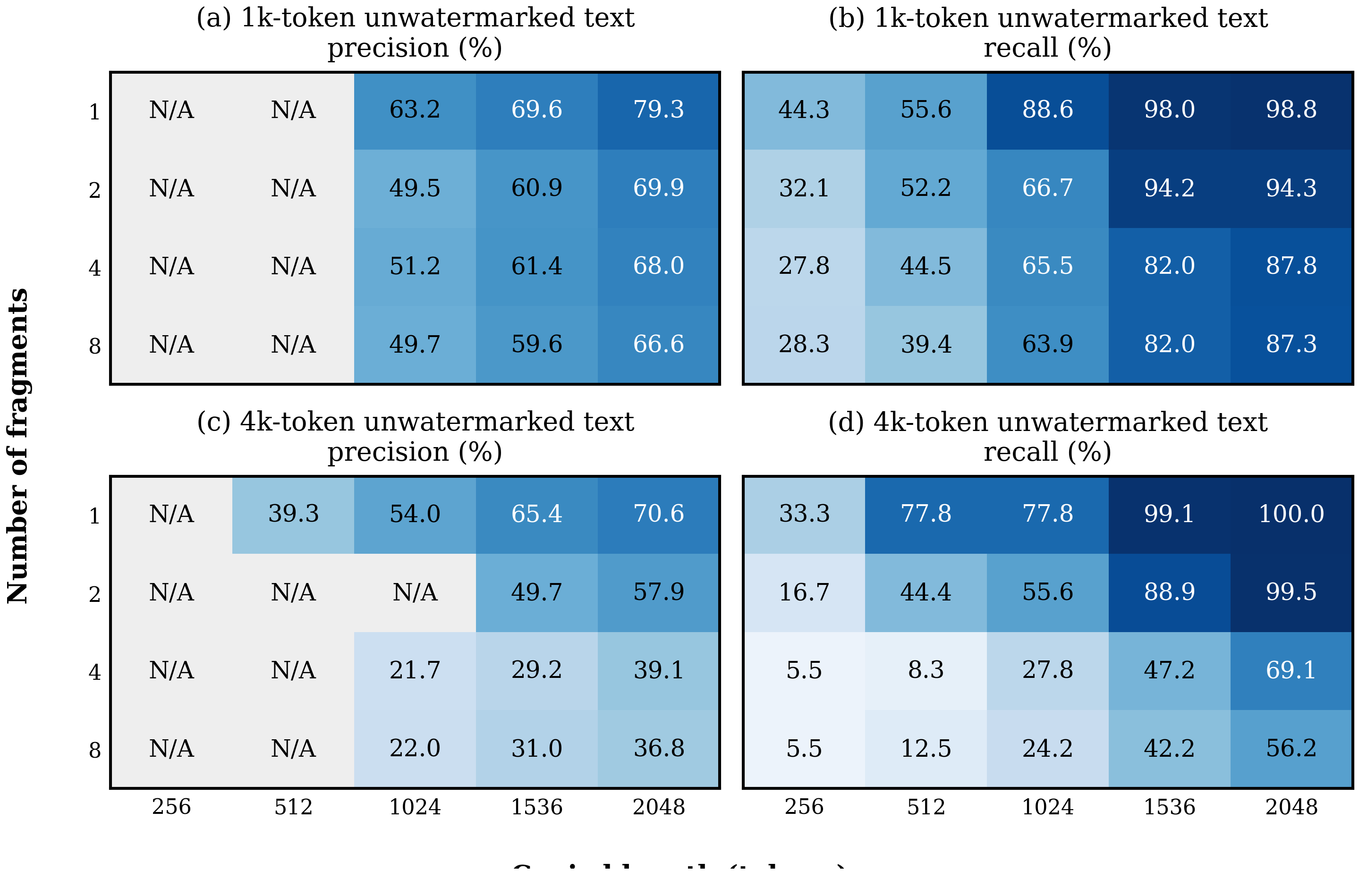}
    \par\smallskip
    {\small\bfseries Copied length (tokens)\par}
    \caption{\textbf{Detection windows also locate the copied content.} In the same copy-paste tests, contiguous 1,536-token passages achieve 98.0--99.1\% localization recall. We report token-level precision and recall as copied length and fragment count vary; scattered fragments are harder to localize.}
    \label{fig:copy_paste_localization}
\end{figure*}

\subsection{Watermark Robustness}
\label{sec:watermark_robustness}

We consider two skill-reuse scenarios: copying watermarked content into an
unwatermarked text (\textbf{copy-paste}) and combining multiple skills (\textbf{skill merging}).
We report detection under both scenarios and localization under copy-paste.

\subsubsection{Copy-Paste}
\label{sec:copy_paste}

\paragraph{Setup.}
We select three skills, from each source, we copy 256, 512, 1024, 1536, or
2048 tokens into an unwatermarked text containing 1024 or 4096 tokens before
insertion. We vary the number of inserted fragments over $\{1,2,4,8\}$.
Within each condition, fragmentation splits the same copied content while
preserving its total length and order.

\paragraph{Detection and localization.}
We scan each document without access to the true insertion
positions, using 1024-token windows with a stride of 256 tokens and a
per-window significance threshold of $p\leq 0.01$.  To evaluate localization,
we merge all significant windows into the predicted region $A$ and take the
union of the inserted fragments as the true region $T$. Overlapping windows count
each token position only once. Localization precision and recall are
\begin{equation}
    \mathrm{Precision}=\frac{|A\cap T|}{|A|},\qquad
    \mathrm{Recall}=\frac{|A\cap T|}{|T|}.
    \label{eq:localization_metrics}
\end{equation}
When no window is significant, recall is zero and precision is undefined.
Each plotted condition includes three seeds for each of the three sources:
we first average within each source and then average the three sources with
equal weight. Precision uses only runs with valid predictions; if any source
has no valid prediction, the corresponding cell is marked N/A. 

\paragraph{Results.}
Figure~\ref{fig:copy_paste_detection} shows that window-based detection consistently identifies copied passages containing as few as 1,536 consecutive tokens. All tested contiguous 1,536-token copies are detected in both 1024- and 4096-token unwatermarked backgrounds. In the longer background, full-text detection reaches only 55.6\%, showing the benefit of window-based detection for identifying copied skill content within unrelated text.
Figure~\ref{fig:copy_paste_localization} shows that the same 1,536-token passages can also be localized, with 98.0--99.1\% recall and 65.4--69.6\% precision across the two background lengths. Thus, detection windows recover nearly all copied content, while also including some surrounding text. Splitting the passage into eight fragments in the 4096-token background reduces detection to 77.8\%, illustrating the effect of dispersed reuse. These results support detecting and locating skill copies for the selected skills with strong watermark signals.

\begin{figure*}[!t]
    \centering
    \includegraphics[width=0.72\textwidth]{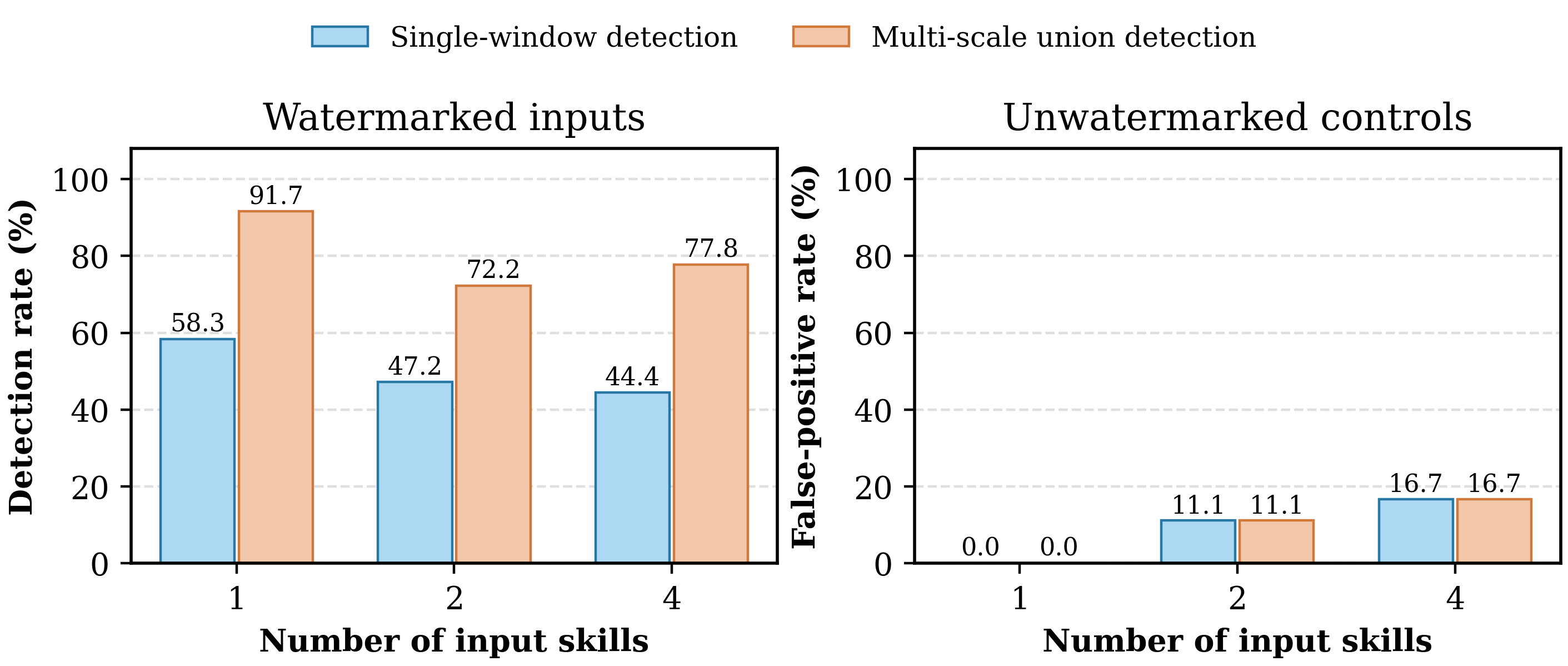}
    \caption{\textbf{Watermarks remain detectable after limited skill merging.} We merge a target skill with zero, one, or three background skills and compare single-window and multi-scale detection. Watermarked targets measure detection rates; unwatermarked targets measure false positives.}
    \label{fig:merge_detection}
\end{figure*}

\subsubsection{Skill Merging}
\label{sec:merge}

Figure~\ref{fig:merge_detection} shows that watermark evidence remains detectable
after merging a target skill with a limited number of background skills.
With one, two, or four total input skills, multi-scale union detection achieves
detection rates of 91.7\%, 72.2\%, and 77.8\%, respectively, compared with
58.3\%, 47.2\%, and 44.4\% for single-window detection. Adding background
skills reduces detection relative to the single-input setting, consistent with
watermark dilution by unwatermarked content. This decline is not strictly
monotonic for multi-scale detection, but both detectors retain lower detection
rates with background skills than without them. Multi-scale detection helps
recover some of the remaining evidence under this dilution.
Both methods have the same observed false-positive rates on unwatermarked
controls: 0.0\%, 11.1\%, and 16.7\% for one, two, and four input skills.
The increasing false-positive rates indicate that reliable detection under
greater background noise also requires calibration.

\subsection{Ablation Study}
\label{sec:ablation}
We ablate five components of SkillMarker: fidelity constraints and logit bias
in quality-constrained rewriting (Section~\ref{sec:quality_rewriting}), and
beam search, signal-oriented selection, and reconstructed-text scoring in
signal-oriented candidate search (Section~\ref{sec:signal_search}). Each variant
removes or replaces one component while retaining literal protection.

\begin{figure}[!ht]
    \centering
    \begingroup
    \includegraphics[width=0.85\columnwidth]{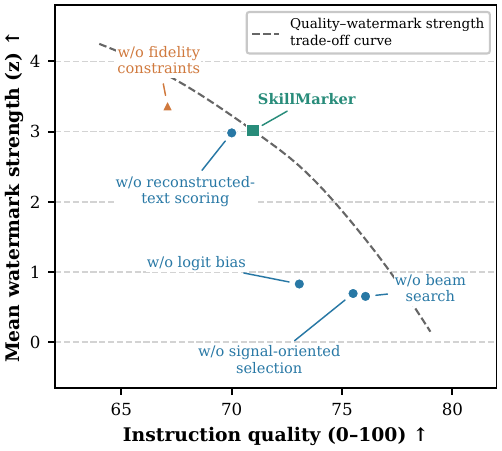}
    \caption{\textbf{Ablation of five components.} We compare instruction quality and watermark strength.}
    \label{fig:ablation}
    \endgroup
    \vspace*{0pt}
\end{figure}
Figure~\ref{fig:ablation} shows the quality--strength trade-off. The KGW $z$-score (Eq.~\eqref{eq:skillmark_detection}) measures the standardized excess of green tokens over the unwatermarked expectation; higher values indicate stronger watermark signals.
SkillMarker achieves quality 70.97 and mean $z=3.01$; removing fidelity
constraints increases $z$ to 3.37 but lowers quality to 67.09.
Removing beam search, signal-oriented selection, or logit bias reduces mean
$z$ to 0.65--0.83 despite higher quality scores, showing their contribution to
watermark evidence. Generated-token scoring remains close to SkillMarker
(70.00 quality, $z=2.98$). Together, these results support balancing instruction
fidelity and watermark strength. Settings, module definitions, and detailed
results are in Appendix~\ref{app:ablation}.

\subsection{Structured Skill Case Study}
\label{sec:structured_case_study}

\begin{figure*}[!t]
    \centering
    \includegraphics[width=0.80\textwidth]{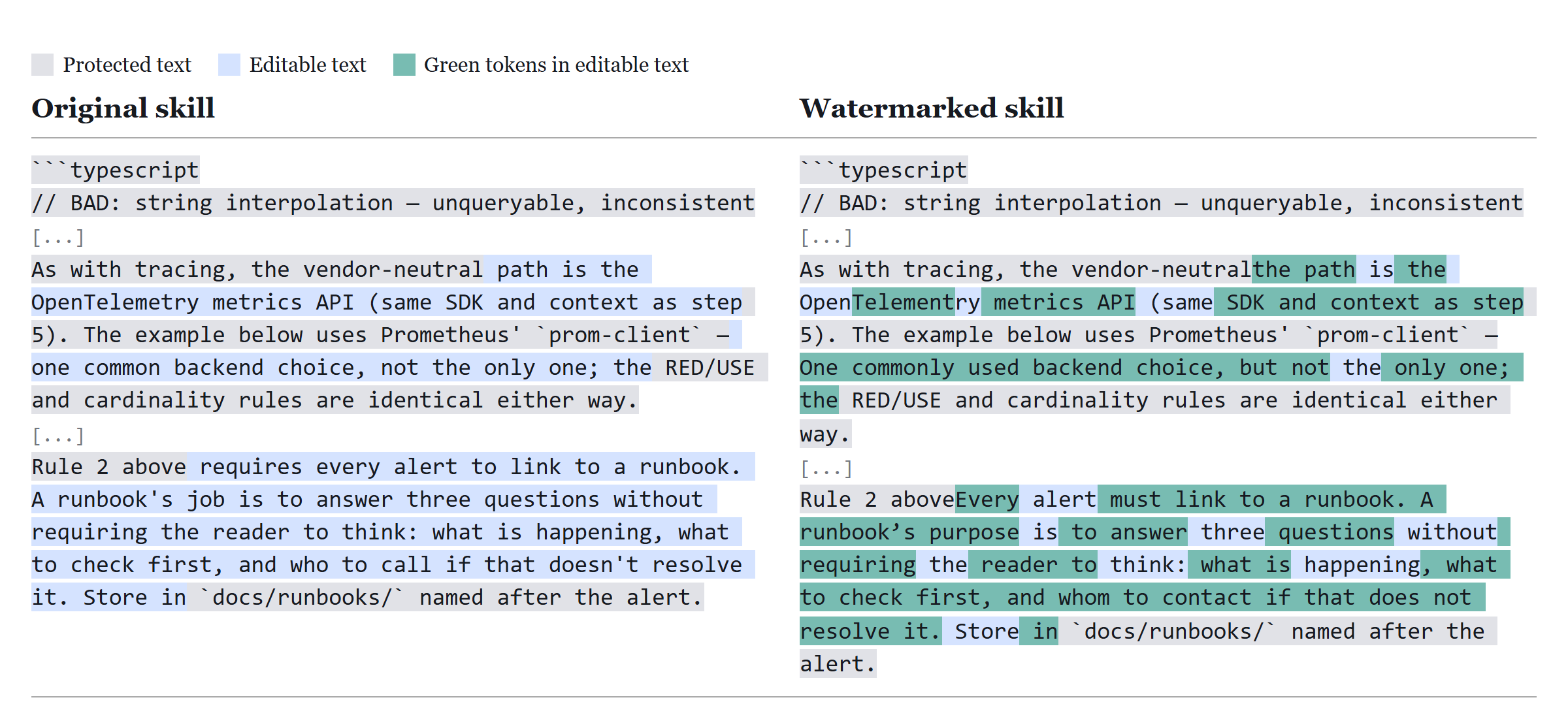}
    \caption{\textbf{Protected content remains unchanged during watermarking.} We compare original and watermarked excerpts of \texttt{observability-and-instrumentation}, showing preserved literal content and rewritten prose that carries the watermark.}
    \label{fig:structured_case_study}
\end{figure*}

Figure~\ref{fig:structured_case_study} compares selected skill excerpts from
\texttt{observability-and-instrumentation} before and after watermarking. Protected spans remain exactly
unchanged, and the saved full text yields nominal $p=4.62\times10^{-9}$.
Long protected blocks show only their opening lines; omission marks separate
non-contiguous excerpts.

%% file: sections/related_work.tex
\section{Related Work}
\label{sec:related_work}

\subsection{Watermarking}
\label{sec:related_watermarking}

\textbf{Generation-time watermarking.} KGW and Unigram-Watermark bias token selection using context-dependent and fixed vocabulary partitions, respectively~\citep{kirchenbauer2023watermark,zhao2024provable}. Other approaches use tournament sampling (SynthID-Text), sentence-level rejection sampling (SemStamp), or optimal transport (textGrain)~\citep{dathathri2024synthid,hou2024semstamp,li2026textgrain}. SWEET and STONE protect low-entropy positions and syntactic code tokens, respectively~\citep{lee2024sweet,kim2026stone}, while AgentMark watermarks planning decisions~\citep{huang2026agentmark}. Generation-time watermarking is typically deployed on the model-provider side to attribute generated outputs. SkillMarker instead enables skill developers to embed watermarks into reusable instructions before distribution, independently of the models that later execute them.

\textbf{Post-hoc watermarking.} PostMark inserts semantically selected words, REMARK-LLM encodes binary signatures, and SrcMarker applies semantics-preserving code transformations~\citep{chang2024postmark,zhang2024remark,yang2024srcmarker}. RedAct watermarks released agent traces through selective rewriting and behavioral signals~\citep{xu2026redact}; SkillTrace audits skill reuse through expression, implementation, and operational traces~\citep{chen2026skilltrace}. These methods establish watermarking and provenance auditing for existing artifacts, but semantic preservation in general text or released traces does not by itself establish instruction fidelity in skills. Skill protection must preserve operational requirements while embedding watermarks that provide sufficient evidence for attribution after reuse. SkillMarker addresses this joint requirement in distributed skill text; reliability studies of copy-paste attacks~\citep{kirchenbauer2024reliability} motivate its evaluation under copy-paste and skill merging.

\subsection{Text Steganography}
\label{sec:related_steganography}

Neural Linguistic Steganography conceals recoverable messages through arithmetic coding~\citep{ziegler2019steganography}; zero-width characters offer invisible encoding but must survive redistribution. ICW induces output patterns through instructions, including covert triggers in modified documents~\citep{liu2026icw}. Its signal is carried by subsequent model outputs, whereas SkillMarker embeds watermarks directly into the distributed skill. Linguistic steganography focuses on concealed message recovery, but its concealment can be compromised by LLM-based steganalysis: fine-tuned LLMs detect steganographic text across tested methods and domains~\citep{bai2024steganalysis,tang2024steganalysis}. Neither objective by itself establishes attribution of copied or merged skill instructions under strict functionality constraints. SkillMarker instead supports ownership verification and leakage tracing through watermark signals under preservation constraints.

%% file: sections/conclusion.tex
\section{Conclusion}
\label{sec:conclusion}

We present SkillMarker, a KGW-based method for protecting distributed agent skills through quality-constrained rewriting and signal-oriented candidate search. Across four datasets, watermarked skills largely preserve downstream performance, with relative changes of $-2.01$\% to $+0.93$\% and dataset-dependent token overhead. Window-based detection improves detection of copied content diluted by unwatermarked text, while fragmentation weakens detection and localization. Watermarks remain detectable after merging with a limited number of background skills, although added background content weakens detection overall and elevated control false positives highlight the need for calibration. These findings support watermarking as a practical direction for skill ownership verification and leakage tracing.

%% file: sections/appendix.tex
\section{Additional Experimental Details}
\label{app:downstream_results}

\subsection{Absolute Downstream Results}

Table~\ref{tab:skillmark_main_results} reports the absolute values underlying
Figure~\ref{fig:downstream_results} for the same settings and datasets.

\input{tables/main_results}

\input{sections/detailed_settings}

\input{sections/experimental_prompts}

\section{Extended Structured Skill Case Study}
\label{app:structured_case_study}

\begin{figure*}[!ht]
    \centering
    \includegraphics[width=\linewidth,height=0.84\textheight,keepaspectratio]{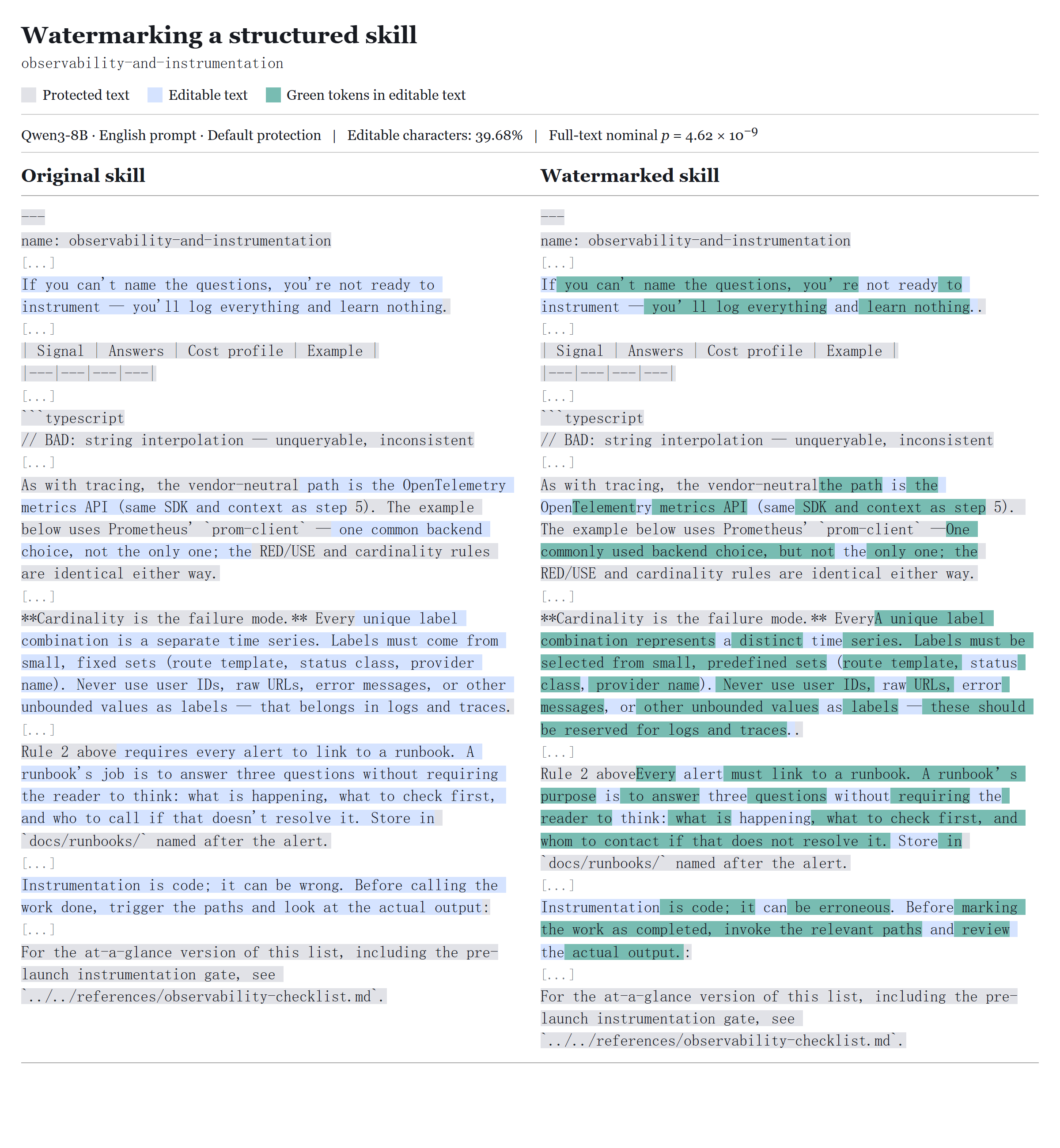}
    \caption{\textbf{Protected content remains unchanged during watermarking.} Extended original and watermarked excerpts of \texttt{observability-and-instrumentation} show preserved literal content and rewritten prose that carries the watermark.}
    \label{fig:structured_case_study_full}
\end{figure*}

Figure~\ref{fig:structured_case_study_full} retains the earlier nine-excerpt
presentation of \texttt{observability-and-instrumentation}, complementing the
compact view in Figure~\ref{fig:structured_case_study}. The skill was watermarked
with Qwen3-8B using an English prompt and default protection. The saved mapping
records exact preservation of protected spans and full-text nominal
$p=4.62\times10^{-9}$. Long protected blocks show their opening lines only;
omission marks separate non-contiguous excerpts. Original generation defects
are retained, so literal preservation and detection do not establish semantic
or functional equivalence.

%% file: tables/main_results.tex
\noindent\begin{minipage}{\textwidth}
    \centering
    \captionof{table}{\textbf{Watermarking largely preserves downstream task performance.} Performance, mean tokens, and median task time across four datasets and three skill settings.}
    \label{tab:skillmark_main_results}
    \footnotesize
    \setlength{\tabcolsep}{1pt}
    \renewcommand{\arraystretch}{1.08}
    \definecolor{watermarkrow}{rgb}{0.91,0.96,0.99}
    \begin{tabular*}{\textwidth}{@{\extracolsep{\fill}}l*{12}{r}@{}}
        \toprule
        & \multicolumn{6}{c}{\textbf{Everyday / General Scenarios}} & \multicolumn{6}{c}{\textbf{Professional Scenarios}} \\
        \cmidrule(lr){2-7}\cmidrule(lr){8-13}
        & \multicolumn{3}{c}{\textbf{agent-skills}} & \multicolumn{3}{c}{\textbf{awesome-skills}} & \multicolumn{3}{c}{\textbf{marketingskills}} & \multicolumn{3}{c}{\textbf{skill-use-bench}} \\
        \cmidrule(lr){2-4}\cmidrule(lr){5-7}\cmidrule(lr){8-10}\cmidrule(lr){11-13}
        \textbf{Setting}
        & \shortstack{Passed\\checks\,$\uparrow$\\(\%)} & \shortstack{Tokens\,$\downarrow$\\(K)} & \shortstack{API\\time\,$\downarrow$\\(s)}
        & \shortstack{Passed\\checks\,$\uparrow$\\(\%)} & \shortstack{Tokens\,$\downarrow$\\(K)} & \shortstack{API\\time\,$\downarrow$\\(s)}
        & \shortstack{Passed\\checks\,$\uparrow$\\(\%)} & \shortstack{Tokens\,$\downarrow$\\(K)} & \shortstack{API\\time\,$\downarrow$\\(s)}
        & \shortstack{Passed\\checks\,$\uparrow$\\(\%)} & \shortstack{Tokens\,$\downarrow$\\(K)} & \shortstack{API\\time\,$\downarrow$\\(s)} \\
        \midrule
        No Skill & 89.83 & 337.5 & 51.6 & 82.93 & 273.1 & 38.2 & 78.10 & 195.0 & 36.9 & 42.66 & 1066.0 & 101.0 \\
        \rowcolor[gray]{0.93}
        Original Skill & 97.46 & 836.6 & 72.2 & 92.03 & 772.1 & 71.2 & 94.20 & 709.8 & 63.2 & 59.85 & 1221.3 & 100.7 \\
        \rowcolor{watermarkrow}
        \textbf{Watermarked Skill} & 98.25 & 954.3 & 82.7 & 90.18 & 998.3 & 75.5 & 93.56 & 879.6 & 73.6 & 60.41 & 1181.3 & 95.4 \\
        \bottomrule
    \end{tabular*}
\end{minipage}

%% file: sections/detailed_settings.tex
\subsection{Main Experimental Hyperparameters}
\label{app:detailed_settings}

The main hyperparameters are listed below. Robustness experiments use the
frozen English watermarked release; its embedding settings are listed separately
from the downstream evaluation.

\begin{center}
\small
\renewcommand{\arraystretch}{1.18}
\begin{tabularx}{\linewidth}{@{}p{0.21\linewidth}p{0.35\linewidth}X@{}}
\toprule
Experiment & Hyperparameter & Value \\
\midrule
Downstream & Agent / judge & DeepSeek-V4-Flash [1m] / same model \\
& Judgments per case & 1 \\
& Bootstrap replicates / seed & 10,000 / 20261010 \\
\midrule
Robustness embedding & Model / beam width & Qwen3-8B / 5 \\
& Green fraction $\gamma$ / logit bias $\delta$ & 0.75 / 8 \\
& Hash key / embedding seed & 15485863 / 123 \\
\midrule
Copy-paste & Copied length (tokens) & 256, 512, 1024, 1536, 2048 \\
& Unwatermarked text length (tokens) & 1024, 4096 \\
& Number of fragments & 1, 2, 4, 8 \\
& Selection / insertion seeds & 123, 456, 789 \\
& Detection window / stride (tokens) & 1024 / 256 \\
& Per-window significance threshold & $p\leq0.01$ \\
\midrule
Skill merging & Merging model / reasoning effort & \texttt{deepseek-flash} / high \\
& Maximum output tokens & 65,536 \\
& Total input skills & 1, 2, 4 (one target plus 0, 1, 3 backgrounds) \\
& Background selection seeds & 123, 456, 789 \\
& Single-window length / stride (tokens) & 3952 / 1027 \\
& Single-window probability threshold & $p\leq0.03332253$ \\
& Multi-scale union & OR of 13 component decisions \\
\bottomrule
\end{tabularx}
\end{center}

Copy-paste window tests use nominal probabilities without multiple-testing
correction. Merge detector parameters were selected on the same evaluation
cohort, without a held-out test set.

%% file: sections/experimental_prompts.tex
\section{Experimental Prompts}
\label{app:experimental_prompts}

\subsection{Watermark Embedding}

The English embedding prompt below contains the system instruction and two
user--assistant demonstrations. The final user message is replaced by the
current editable passage. Role labels and the bracketed input field are
presentation annotations; the remaining text is reproduced verbatim from the
saved generation configuration.

\begin{appendixpromptbox}{Watermark embedding: English prompt with demonstrations}
\begin{appendixprompttext}
[System]
Preserve the original meaning, key facts, conditions, negations, numerical values, technical identifiers, and formatting. You may freely rephrase the remaining wording. Output only the rewritten passage in its original language.

[User: demonstration 1]
**Core principle**: Automatically write patterns that meet the automatic-write conditions (frequency ≥3, consistent fixes, certainty 100

[Assistant: demonstration 1]
**Core principle**: Write patterns to skill references automatically only when all automatic-write conditions are met (frequency ≥3, consistent fixes, certainty 100

[User: demonstration 2]
Purely visual ALIGN differences in CHECK-8 and CHECK-11 (style compliance) do not block the gate; record them in the report for later improvement. **CRASH/URL issues in CHECK-8 are functional failures** (a page failing to open or crashing is not a "visual issue"); they block the gate and enter the §5 loop.

[Assistant: demonstration 2]
Record purely visual ALIGN differences in CHECK-8 and CHECK-11 (style compliance) in the report for future improvement; they do not block the gate. **CRASH/URL issues in CHECK-8 are functional failures** (a page failing to open or crashing is not a "visual issue"), so they block the gate and enter the §5 loop.

[User: actual input]
[CURRENT EDITABLE PASSAGE]
\end{appendixprompttext}
\end{appendixpromptbox}

\subsection{Skill-Merging Instruction}

For the merging evaluation, the system message contains the instruction below;
the user message contains the complete input skills.

\begin{appendixpromptbox}{Skill-merging evaluation prompt}
\begin{appendixprompttext}
[System]
Merge the provided skills into one skill.

[User]
[COMPLETE INPUT SKILLS IN THE RECORDED ORDER]
\end{appendixprompttext}
\end{appendixpromptbox}

%% file: sections/arr/ablation.tex
\section{Ablation Settings and Additional Results}
\label{app:ablation}

\paragraph{Experimental setup.}
We evaluate 32 selected skills from the four datasets using Qwen3-8B \citep{yang2025qwen3}, with three
fixed seeds per skill and configuration. SkillMarker uses five-beam search,
$\gamma=0.75$, and logit bias $\delta=8$. We average the three runs within each
skill and then weight skills equally. Instruction quality is a blind, holistic
model-review score on a 0--100 scale, rather than an execution-based measure of
functional correctness. Watermark strength is the full-document $z$-score after
saving and retokenizing the output. Figure~\ref{fig:ablation} reports means;
an average $z$-score is not a document-level significance test or detection rate.

\paragraph{Module correspondence.}
All six configurations retain mandatory literal protection.
For quality-constrained rewriting (Section~\ref{sec:quality_rewriting}),
\emph{w/o fidelity constraints} removes the prompt-level preservation
constraints, while \emph{w/o logit bias} sets $\delta=0$ but retains candidate
selection. For signal-oriented candidate search
(Section~\ref{sec:signal_search}), \emph{w/o beam search} uses a single path;
\emph{w/o signal-oriented selection} replaces watermark-oriented ranking with
model-score selection; and \emph{w/o reconstructed-text scoring} scores generated
tokens instead of the reconstructed, retokenized document. The latter two
variants replay SkillMarker's candidate pools, isolating the selection rule.
Reconstructed-text scoring is distinct from token-variant grouping, which remains
enabled.

\paragraph{Results.}
SkillMarker achieves quality 70.97 and mean strength 3.01.
Removing fidelity constraints raises strength to 3.37 but lowers quality to
67.09, illustrating the role of these constraints in preserving instructions.
Removing beam search, signal-oriented selection, or logit bias reduces strength
to 0.65, 0.69, or 0.83, respectively, while quality increases to 76.06, 75.50,
or 73.06. These components therefore contribute watermark evidence at a cost
in holistic quality score. Generated-token scoring yields 70.00 quality and
2.98 strength, close to SkillMarker; these means alone do not establish
a substantial benefit from reconstructed-text scoring. Overall, SkillMarker
offers a balance rather than dominating every alternative. The curve is a
visual guide anchored at SkillMarker, not evidence of attainable intermediate
configurations.